\documentstyle[epsfig,aps,prl,bbm,twocolumn,doublespace,12pt]{revtex}
\renewcommand{\theequation}{\arabic{equation}}
\begin{document}
%
%
%
%
\onecolumn
\title{
On the limiting procedure by which $SDiff(T^2)$ and
$SU(\infty)$ are associated}
\author{John Swain}
\address{Department of Physics, Northeastern University, Boston, MA 02115, USA\\
email: john.swain@cern.ch}
\date{April 29, 2004}
\maketitle

\begin{abstract}
\section*{\bf Abstract}

There have been various attempts to identify groups of area-preserving
diffeomorphisms of 2-dimensional manifolds with limits of $SU(N)$ 
as $N\rightarrow\infty$. We discuss the particularly simple case
where the manifold concerned is the two-dimensional torus $T^2$
and argue that the limit, even in the basis commonly used, is ill-behaved
and that the large-N limit of $SU(N)$ is much larger than $SDiff(T^2)$.
\end{abstract}

\section{Introduction}

Groups of area-preserving diffeomorphisms and their Lie algebras
have recently been the focus of much attention in the physics 
literature.    
Hoppe \cite{Hoppe1} has shown that in a suitable basis, the Lie algebra of 
the group $SDiff(S^2)$ of area-preserving diffeomorphisms of a
sphere tends to that
of $SU(N)$ as $N\rightarrow\infty$.
Similar arguments have been made associating various infinite limits of
Lie algebras of classical groups with Lie algebras of groups of
area-preserving diffeomorphisms of 2-dimensional surfaces.
This has obvious interest in 
connection with gauge theories of $SU(N)$ for large N. 
The use of $SU(N)$ for finite $N$ as
an approximation to groups of area-preserving diffeomorphisms has also been
used in studies of supermembranes \cite{supermem1,supermem2,supermem3} and 
in particular has been used to argue for their instability.
The authors of references \cite{supermem2} and \cite{supermem3}
have especially
emphasized the difficulties in relating such infinite limits with Lie algebras
of area-preserving diffeomorphisms.
Various authors
have considered special limits and/or large-N limits of
other classical Lie algebras \cite{Pope,Pope2,Wolski,deWit,Vassilevich}
as relevant for 2-manifolds other than spheres. 
The purpose of this Letter is to clarify the nature of the
limiting procedure by which $SU(\infty)$ has been related to $SDiff(T^2)$.

\section{The Lie algebras of $SDiff(T^2)$}

We follow here the treatment of \cite{Pope2}, which is particularly clear.
The torus $T^2$ is represented by the plane ${\mathbbm{R}}^2$ with
coordinates $x$ and $y$ and the identifications 

\begin{equation}
(x,y) = (x + 2\pi,y)
\end{equation}

\noindent and

\begin{equation}
(x,y) = (x,y + 2\pi)
\end{equation}

A basis for functions on the torus is chosen as

\begin{equation}
Y_{mn}(x,y) = exp\left[i(mx+ny)\right]
\end{equation}
 
\noindent with $m,n$ running over all integers. The local area-preserving
diffeomorphisms are then generated by the vector fields

\begin{equation}
L_{mn} = ( \epsilon^{ab}\partial_b Y_{mn} ) \partial_a = 
i exp \left[ i(mx+ny)\right]( n\partial_x - m \partial_y )
\end{equation}

\noindent with indices $a,b=1,2$. In other words, the
divergence-free vector fields are those which are the 
curl of something else.

The generators clearly close under commutation, with the commutator

\begin{equation}
\left[ L_{mn}, L_{m^\prime,n^\prime} \right] = (m n^\prime - m^\prime n) 
L_{m+m^\prime,n+n^\prime}
\end{equation}
\label{eq:sdiff}

\section{The Lie algebra of $SU(N)$}

To construct the Lie algebra of $SU(N)$, again following \cite{Pope2},
we sketch the basic idea.
Fix a positive integer $N$ and a complex number $\omega$ such 
that $\omega^N=1$ but $\omega^r\neq 1$ for $0<r<N$. 
$\omega$ is called a primitive root of unity.
Then we have
$\omega = exp(2\pi i k/N)$ for some $k$ relatively prime to $N$.
Now we find unitary, traceless matrices $g$ and $h$ 
such that

\begin{equation}
hg = \omega g h
\end{equation} 

Then the set of matrices

\begin{equation}
J_{m,n} = \omega^{mn/2}g^m h^n
\end{equation}
\label{eq:defJ}

\noindent for $0 \leq m,n < N$ are linearly independent and are a basis for
the $N\times N$ matrices. $J_{0,0} = 1$, and all the other 
$J_{m,n}$ are traceless and satisfy $J_{m,n}^\dag = J_{-m,-n}$.
Leaving out $J_{0,0}$, the scaled matrices $J^\prime_{m,n} = iN/(2k\pi)J_{m,n}$
generate $SU(N)$ with the commutation relations 

\begin{equation}
\left[ J^\prime_{m,n}, J^\prime_{m^\prime,n^\prime}\right] = 
\frac{N}{k\pi} \sin\left( \frac{k\pi}{N}(m n^\prime - m^\prime n) \right)
J^\prime_{m+m^\prime,n+n^\prime}
\end{equation}
\label{eq:sun}

\section{The $N\rightarrow\infty$ Limit}

The claim now is that in the limit $N\rightarrow\infty$ that the
commutation relations in equation \ref{eq:sun} go over to those in
equation \ref{eq:sdiff}. Naively, of course, one would like to 
argue that as $N\rightarrow\infty$,

\begin{equation}
\frac{N}{k\pi}\sin\left(\frac{k\pi}{N}(m n^\prime - m^\prime n)\right) = 
(m n^\prime - m^\prime n) + O(1/N^2)
\end{equation}

\noindent and drop the terms of order $1/N^2$ and higher. However, let us keep
the next term and examine whether or not it can indeed be taken to be
small.

\begin{equation}
\frac{N}{k\pi}\sin\left(\frac{k\pi}{N}(m n^\prime - m^\prime n)\right) = 
(m n^\prime - m^\prime n) -\frac{1}{3!} \frac{(k\pi)^2}{N^2}(m n^\prime - m^\prime n)^3
+ \ldots
\end{equation}

Now consider any choice of $(m,n) = (N/a,0)$ and $(m^\prime,n^\prime)=(0,N/b)$
where $a$ and $b$ are arbitrary integers that divide $N$ (including one).
Then 

\begin{equation}
\frac{(k\pi)^2}{N^2}(m n^\prime - m^\prime n)^3 = 
\frac{(k\pi)^2}{a^3 b^3} N^4 
\end{equation}

\noindent which is clearly {\it not} negligible as $N\rightarrow\infty$.
It would seem that there are many elements of the Lie algebra of
$SU(N)$ which do not belong to $SDiff(T^2)$.

This is in keeping with ideas raised in 
\cite{myself} suggesting that $SU(\infty)$ is much larger
than the group of area-preserving diffeomorphisms of a surface,
and perhaps descibes some sort of theory incluing topology change.
Other work demonstrating that topologically, 
$SDiff(T^2)$, and indeed all the area-preserving diffeomorphism
groups, are inequivalent to $SU(\infty)$ is in \cite{topology}.

\section{Acknowledgement}

The author would like to
my colleagues at Northeastern University and the National Science
Foundation for their support.

\end{document}